\documentstyle[11pt,paspconf]{article} 
\begin{document}

\title{Physical processes in pulsar magnetospheres and
non-thermal high-frequency emission }  

\author{Vladimir V. Usov } 
\affil{Department of Physics, Weizmann Institute, Rehovot 76100,
Israel}  

\begin{abstract}
The energy source of a pulsar's non-thermal emission is the rotational
energy loss of the neutron star. The rotational energy of the neutron
star is transformed into the pulsar radiation by a long
sequence of processes. The processes of this sequence are discussed.
\end{abstract}


\section{ Introduction }

The current sample of radio pulsars contains about six hundred
pulsars (e.g., Taylor {\it et al.} 1993). The radio luminosities of the
pulsars are small  compared with the energy loss rate due to the pulsar
spin down ($\sim 10^{-6}-10^{-5}$). Strong high-frequency radiation
in the optical, X-ray and $\gamma$-ray ranges is observed from a few
radio pulsars (for a review, see
Thompson 1994; Hartmann 1995). The observed radiated power for these
pulsars is concentrated mainly in the $\gamma$-ray range. The
$\gamma$-ray pulsar Geminga is probably also a radio pulsar
(Halpern \& Holt 1992) which is "radio quiet" because its radio
beam does not intersect the Earth (Ozernoy \& Usov 1977). The
pulsar luminosities in $\gamma$-rays are a substantial fraction
($\sim 10^{-3}-10^{-1}$) of the spin-down power, $\dot E_{\rm rot}$.
This makes studies of
high-energy radiation of pulsars a promising avenue to
a better understanding of physical processes which result in
non-thermal radiation of pulsars.
\par
A common point of all available models of pulsars is that a strong
electric field is generated in the magnetosphere of a rotating
magnetized
neutron star (e.g., Michel 1991). The component of the electric
field $E_\parallel =({\bf E\cdot B})/|{\bf B}|$ along
the magnetic field ${\bf B}$ is non-zero, and this $E_\parallel$
can accelerate particles to ultrarelativistic energies.
The accelerated particles emit $\gamma$-rays due to curvature
emission and other processes. Some of these $\gamma$-rays are absorbed
by creating secondary electron-positron pairs. The created
electron-positron pairs  screen the electric field $E_\parallel$
in the pulsar magnetosphere everywhere except for the compact regions.
The regions where $E_\parallel$ is unscreened are called gaps.
These gaps are, in fact, an "engine" which is responsible for
the pulsar radiation.
\par
Two kinds of gap models have been popular in trying to
explain the non-thermal radiation of pulsars. The main difference
between these two is the site of gaps. A gap that forms
near the magnetic poles of the pulsar is called a polar gap.
Besides polar-gap models there are also outer-gap models (e.g.,
Cheng {\it et al.} 1986 a,b; Chiang \& Romani 1994;
Usov 1994; Romani and Yadigaro\v glu 1995).
This review mainly concentrates on the polar gap models (about
outer gaps, see Romani, these proceedings).

\section{ The polar gap models}

Polar-gap models for pulsars may be classified in two ways:
whether ions or electrons tend to be ejected from the surface, and
whether $E_\parallel$ is zero or non-zero at the stellar surface.
The latter depends on the character of the particle outflow
from the surface.
\par
The kind of the particles that tend to be ejected from the
neutron star surface by the field $E_\parallel$
depends on the sign of ${\bf \Omega \cdot B}$, where ${\bf \Omega}$
is the angular velocity of the pulsar rotation.
Electrons tend to be ejected for ${\bf\Omega\cdot B}>0$ and ions for
${\bf\Omega\cdot B}<0$.

The most familiar model in which there is no ejection of particles
from
the stellar surface is that of Ruderman \& Sutherland (1975).
In this model, the field $E_\parallel$ is maximal at the surface and
decreases with distance.
In contrast, in the model of Arons (1981) it is assumed that
charged particles flow freely from the neutron star surface. In this
model the electric field $E_\parallel$ is equal to zero at the surface
and increases with distance above the surface.
A third kind of polar-gap model (Cheng \& Ruderman 1980; Usov \&
Melrose 1995, 1996)
is an intermediate case where the particles flow from the pulsar
surface but not freely. In such a model the field $E_\parallel$ is
non-zero
at the pulsar surface, but it is smaller than in the
model of Ruderman \& Sutherland (1975).

\subsection{ The surface structure and particle ejection}

The structure of matter in the surface layers of neutron stars with
$B_{_S}\gg\alpha^2 B_{\rm cr}\simeq 2.35\times10^9$ G
is largely determined by the magnetic field. The neutron star surface
may be solid provided that the
surface temperature is smaller than the melting temperature
(e.g., Liberman \& Johansson 1995;
Usov \& Melrose 1995), here $B_{\rm cr}=m^2c^3/e\hbar =4.4\times
10^{13}$ G, $\alpha=e^2/\hbar c=1/137$ is the fine structure
constant. In this case charges can escape from the surface due
to thermionic emission. One
may define a characteristic temperature, $T_e$ (for ${\bf\Omega\cdot
B}>0$) or $T_i$ (for ${\bf\Omega\cdot B}<0$), such that for a
surface temperature $T_{_S}<T_e$ or $T_{_S}<T_i$ thermionic emission
is negligible. For a magnetic metal with iron ions and
$B\simeq 10^{12}-10^{13}$ G the characteristic temperature
is $T_e\simeq 4\times 10^5(B_{_S}/10^{12}\,{\rm G})^{0.4}$ K for
electrons and
$T_i\simeq 3.5\times 10^5(B_{_S}/10^{12}\,{\rm G})^{0.73}$ K
for ions within a factor of 2 (e.g., Usov \& Melrose 1995).

The flow of charged particles away from the solid surface is
very sensitive to the surface temperature for $T_{_S}\sim T_e$
(or $T_{_S}\sim T_i$).
A small change in $T_{_S}$ around $T_{_S}\sim T_e$ (or $T_{_S}\sim
T_i$) can have a large effect on the density, $n_e$ (or $n_i$),
of outflowing
particles, with a change by factor of two causing $n_e/n_{_{\rm GJ}}$
(or $n_iZ/n_{_{\rm GJ}}$)
to vary from exponentially small (the Ruderman-Sutherland model)
to unity (the Arons model), where $n_{_{\rm GJ}}
={\vert {\bf \Omega \cdot B}\vert /2\pi ce}$ and $Z$ is the ion charge.

\subsection{Acceleration of outflowing particles}

The Goldreih-Julian density $n_{_{\rm GJ}}$ is determined so that
the electric field $E_\parallel$ in the outflowing plasma
is screened completely
if the charge density is equal to $en_{_{\rm GJ}}$. Therefore,
the accelerating field $E_\parallel$ arises only
due to deviations from this density. Many causes can lead to such
deviations. They are (1) the inertia of particles (Michel 1974),
(2) the curvature of the magnetic field lines (Arons 1981),
(3) the General Relativity effects (Muslimov \& Tsygan 1992), and
(4) the binding of particles within the neutron star surface
(Ruderman \& Sutherland 1975; Cheng \& Ruderman 1980; Usov \&
Melrose 1995, 1996).
\par
When the neutron star surface is cold enough (see 2.1) and
there is no ejection of
particles from the surface, the binding of particles
determines the $E_\parallel$-field distribution in the pulsar
vicinity. In the case
when particles flow freely from the stellar surface,
both the generation of the field $E_\parallel$
and acceleration of particles in the polar gaps are mainly
because of the General Relativity effects.

\subsection{Generation of $\gamma$-rays}

In a strong magnetic field near the pulsar surface, electrons (and
positrons) lose the momentum component transverse to the magnetic field
very rapidly and move away from the pulsar
practically along
the field lines. For such electrons in the ground-state Landau level,
any energy loss is negligible up to the Lorentz factors of $\sim 10$.
For $10\la \Gamma\la 10^2$, the energy loss due to cyclotron
resonant scattering of thermal X-rays from the neutron star surface
increases sharply (Dermer 1990). The mean
energy of scattered photons is $\sim \hbar \omega_{_B}\Gamma\simeq
(B/10^{12}\,{\rm G})(\Gamma /10^5)$ GeV, where $\omega _{_B}=
eB/mc$ and $\Gamma$ is the electron
Lorentz factor.
\par
Magnetic Compton scattering is the dominant energy-loss process
near the neutron star surface
when the electron Lorentz factors are less than $\sim 10^6$
for typical $\gamma$-ray pulsar magnetic fields and surface
temperatures measured by ROSAT (e.g., Sturner 1995).
At $\Gamma > 10^6$, the main energy loss for
ultrarelativistic electrons in the pulsar magnetospheres is due
to curvature radiation. In this case the rate of energy loss is
$|{\dot\varepsilon}_e|= {2e^2c/3 R_{\rm c}^2}\Gamma^4$, and
the mean energy of curvature photons is
$\bar\varepsilon_\gamma= {3\hbar c/2 R_{\rm c}}\Gamma^3$
(e.g., Ochelkov \& Usov 1980), where $R_{\rm c}$ is the radius of
the curvature of the magnetic field lines.
\par
For all known pulsars, $\gamma$-rays generated near the neutron
star surface are produced in a state below the pair creation
threshold (Usov \& Melrose 1995).

\subsection{Propagation of $\gamma$-rays and pair creation}

The conventional expression for the refractive index of plasma,
with the vacuum polarization by the magnetic field taken into
account, differs from unity by the order of
$\lbrack0.1\alpha(B/B_{\rm cr})^2
+(\omega_p/\omega)^2\rbrack
\sin^2\vartheta$, with
$\omega_p=(4\pi e^2n_p/m)^{1/2}$,
where $n_p$ is the plasma density, $\omega=\varepsilon_\gamma/\hbar$
is the
photon frequency and $\vartheta$ is the angle between the photon wave
vector ${\bf k}$ and the magnetic field ${\bf B}$ (Erber 1966; Adler
1971). For $\omega\gg3\alpha^{-1/2}(B_{\rm cr}/B)\omega_p$, the
vacuum polarization is the main contributor to the difference between
the refractive index and unity. This condition
is well satisfied for $\gamma$-rays near the pulsar surface.
Hence, to understand the process of $\gamma$-ray
propagation in the vicinity of pulsars it
suffices to consider propagation in the vacuum polarized by a strong
magnetic field.

\par The principal modes of propagation for a photon in the
magnetized
vacuum are linearly polarized with electric vectors either
perpendicular ($\perp$ mode) or parallel ($\parallel$ mode) to the
plane formed by the photon wave vector ${\bf k}$ and the vector
${\bf B}$ (e.g., Adler 1971).

While the photon is below the pair creation threshold,
$\varepsilon_\gamma\sin\vartheta<2mc^2$ for $\parallel$ mode and
$\varepsilon_\gamma\sin\vartheta<2mc^2\{1+[1+(2B/B_{\rm cr})]^{1/2}
\}$ for $\perp$ mode,
its main (inelastic) interaction with the magnetic field is a
splitting into two photons $\gamma + B \rightarrow \gamma'+\gamma''+
B$ (Adler 1971; Usov \& Shabad 1983; Baring 1991
and references therein). Under the assumption that the dispersion is
small, Adler (1971) showed that only $\perp$ mode may undergo the
decay, $\perp\to \parallel+\parallel$. Towards the threshold,
however,
the dispersion law differs considerably from the vacuum case,
$\omega =|{\bf k}|c$, and the small-dispersion assumption is no
longer applicable. The decay conditions
for the resonant dispersion law were studied by Usov \& Shabad (1983)
who showed that Adler's conclusion remains unaffected by taking
the resonant effects into consideration.

The coefficient of photon absorption by decay in the weak-field limit
$B\ll B_{\rm cr}$ is $\sim 0.1 (B\sin\vartheta /B_{\rm cr})^6
(\varepsilon_\gamma /mc^2)^5$ cm$^{-1}$. Recently, it was claimed
(Wunner {\it et al.} 1995) that this formula underestimates the
correct splitting rate by several orders of magnitude at $B\sim
(0.1-1)B_{\rm cr}$. However, the paper of Wunner {\it et al.} (1995)
was criticized by Baier {\it et al.} (1996).

If the strength of the magnetic field at the pulsar poles is high
enough,
$B_{_S}\ga 0.2\,B_{\rm cr}$, most of the $\perp$-polarized photons
with $\varepsilon_\gamma\la 10^2\rm\,MeV$ produced near
the pulsar surface, are split and transformed into
$\parallel$-polarized
photons before the pair creation threshold is reached (Usov \&
Shabad 1983; Usov \& Melrose 1995). As a result, the
$\gamma$-ray emission recorded from the pulsar vicinity at energies
$\varepsilon_\gamma\la10^2\rm\,MeV$ may be linearly polarized up to
100 \%. By observing the polarization of the $\gamma$-ray emission of
pulsars it would be possible to estimate the strength of the magnetic
field near the pulsar surface.

If the photon energy is above the pair creation threshold, the
main process by which a photon interacts with the magnetic field is
single-photon absorption, accompanied by pair creation:
$\gamma+B\rightarrow e^++e^-+B$ (Erber 1966;
Adler 1971; Baring 1991). In the application to pulsars
it is usually assumed that the $\gamma$-quanta produced by the
accelerated electrons below the pair creation threshold propagate
through the pulsar magnetosphere until they are absorbed by creating
free pairs. However, before a photon reaches the threshold for free
pair creation it must cross the threshold for bound pair creation.
The assumption that the created
pairs are free is not valid if the magnetic field is strong
enough, specifically for $B>0.1\,B_{\rm cr}$.
In such a strong magnetic field, the $\gamma$-quanta
emitted tangentially to the curved force lines of the magnetic field
are captured near the threshold of bound pair creation and are then
channelled along the magnetic field as positronium, that is, as
bound pairs (Shabad \& Usov 1985, 1986; Herold {\it et al.} 1985).
This positronium may be stable in the polar gaps against both the
ionizing action of the electric field and against photo-ionization
(Shabad \& Usov 1985; Bhatia {\it et al.} 1992;
Usov \& Melrose 1995).

The fact that for $B_{_S}>0.1B_{\rm cr}$ the electron-positron
pairs created in the neutron star vicinity are bound may be
very important for many physical processes in the pulsar
magnetosphere. For example, positronium atoms form a gas of
electroneutral particles.
Such a gas does not undergo plasma processes, like plasma
instabilities, which are responsible for generation of the
pulsar radio emission. Maybe, the suppression of free pair
creation in strong magnetic fields results in a
death line of pulsars at $B_{_S} \sim 10^{13}$ G
(Arons 1995, private communication). Besides,
unlike free pairs, such bound pairs do not screen the electric field
$E_\parallel$ near the pulsar. Screening requires a net charge
density which can build up due to free pairs being separated by
$E_\parallel$,
but cannot build up if the pairs remain bound. As a result the pulsar
luminosity is higher than it would have been in the absence of
formation of positronium (Shabad \& Usov 1985; Usov \&
Melrose 1995, 1996 and below).

\subsection{Non-thermal luminosities}

The total power carried away by both relativistic particles and
radiation from the polar gap into the pulsar magnetosphere is
$$L_p\simeq
\dot N_{\rm prim}e\Delta\varphi \,,
\eqno(1)$$

\noindent where $\dot N_{\rm prim}$ is the flux of primary electrons
(or positrons) from the polar cap
and $\Delta\varphi$ is the potential across the polar gap.
Equation (1) is valid regardless of whether the electron-positron
pairs created near the pulsar are free or bound. The version of
pair creation determines only the $\Delta \varphi$ value.

Ruderman \& Sutherland (1975) were the first ones to develop a
self-consistent polar-gap model in which the screening of the
field $E_\parallel$ by the pairs created
in it is taken into account. Consideration of this screening led
Ruderman \& Sutherland (1975) to conclude that the potential across
the polar gap cannot exceed $\Delta\varphi_{_{\rm RS}}\simeq$ a few
$\times10^{12}\rm\,V$. This upper limit on $\Delta \varphi$
is valid for any polar-gap model in which free pairs are
created by $\gamma$-quanta absorption in the magnetic field.

The density of the primaries cannot be more than $n_{_{\rm GJ}}$.
Therefore, we have the following upper limit
on the flux of the primaries, $\dot N_{\rm prim}\leq n_{_{\rm GJ}}c
\Delta S$, where $\Delta S$ is the surface of the polar cap.

It is convenient to define the ratio $\eta_\gamma =
L_p/\dot E_{\rm rot}$ of the spin-down power going into both
high-energy particles and radiation.
For $\dot N_{\rm prim}= n_{_{\rm GJ}}c \Delta S$ and $\Delta \varphi=
\Delta \varphi_{_{\rm RS}}$ the corresponding fraction is
(e.g., Usov \& Melrose 1995)

$$\eta_\gamma^f\simeq
1.5\times10^{-3}
\left({B_{_S}\over 0.1\,B_{\rm cr}}\right)^{-8/7}
\left({P\over 0.1\,{\rm s}}\right)^{15/7}.
\eqno(2)$$

\noindent
where $P=2\pi /\Omega$ is the pulsar period.

In conventional polar-gap models (e.g., Ruderman \& Sutherland 1975;
Arons 1981; Cheng \& Ruderman 1980) where created pairs are
free, the value of $\eta_\gamma^f$ is more or less the same and
differs
from (2) by a factor of 2 or so. From Table 1 we can see that
$\eta^f_\gamma$ is more than an order of magnitude smaller than
$\eta_\gamma^{\rm obs}=L_{X+\gamma}/\dot E_{\rm rot}$, i.e.
the inferred high efficiency of conversion of
rotational energy into $\gamma$-ray radiation
cannot be explained within the framework of these models.

\begin{table}
\caption{Properties of $\gamma$-ray pulsars}
\begin{center}

$$\vbox{
\tabskip=0pt
\halign{
\hfil#\hfil&
\hfil$\;#\;$\hfil&
\hfil$\;#\;$\hfil&
\hfil$\;#\;$\hfil&
\hfil$\;#\;$\hfil&
\hfil$\;#\;$\hfil&
\hfil$\;#\;$\hfil&
\hfil$\;#\;$\hfil&
\hfil$\;#\;$\hfil
\cr
\noalign{\hrule\vskip5pt}
Name   & P  &   B_{_S}
   &   D   &      L_{X+\gamma}&
\dot E_{\rm rot}  &  \eta^{\rm obs}_\gamma&        \eta^f_\gamma &
\eta^b_\gamma \cr
\noalign{\vskip3pt}
&{\rm ms}&10^{12}{\rm G}&{\rm kpc}&10^{36}{\rm erg\, s^{-1}}&
10^{36}{\rm erg\, s^{-1}} &10^{-2}&10^{-2}&10^{-2} \cr
\noalign{\vskip5pt\hrule\vskip5pt}
PSR 0531+21 &   33  &  6.6   &   2    &    2.2&
450 &  0.5    &     0.01  &  1.7\cr
PSR 0540--69 &   50   &   9  &   55   & 0.9 &  150
&  0.6 &  0.02 & 2.3\cr
PSR 0833--45  &  89  &  6.8 &   0.5  &  0.084&   7
 & 1.2 & 0.08 & 4.6\cr
PSR 1706--44  & 102  &  6.3 &   2.8     & 0.3  &  3.4
&  9  &  0.1 &    6\cr
PSR 1509--58  & 150  &  31  &  4.2  &0.39&  20
& 2  & 0.04 & 3.6\cr
PSR 1055--52 &  197  &  2 & 1.8& 0.024& 0.03 &
80  &  1 & -\cr
&& (6) & & & & & & 22\cr
Geminga   &    237  &  3.3 & (0.15)
& 0.003&0.035& 9  & 1&-\cr
&& (6) & & & & & & 27\cr
\noalign{\vskip5pt\hrule}
}
}$$

\end{center}
\end{table}

One suggestion to explain the contradiction between the polar-gap
theory and the $\gamma$-ray observations
is that the rotation and magnetic axes of the $\gamma$-ray pulsars
are nearly aligned, and the $\gamma$-ray radiation
is strongly beamed (Dermer \& Sturner 1994; Daugherty \& Harding
1994).
However, these small beam widths imply that the
chance of observing any given pulsar from the Earth is too
small, about $10^{-2}$ (Daugherty \& Harding 1994)

Recently, Usov \& Melrose (1995, 1996) developed the
modified polar-gap model that involves a greater power
going into primary particles than conventional models,
if the production of free
pairs is suppressed, as occurs in a sufficiently strong magnetic
field, $B>0.1B_{\rm cr}$. In this model,
the fraction of the spin-down power going into both
high-energy particles and radiation is

$$\eta^b_\gamma\simeq
{3\over2}\left({P\over P_1}\right)^{3/2}
\left[1-\left({P\over P_1}\right)^{3/2}\right]\,,
\,\,\,\,\,{\rm where}\,\,\,\,\,
P_1\simeq 0.5
\left({B_{_S}\over 0.1\,B_{\rm
cr}}\right)^{2/3} \,{\rm s}
\,. \eqno(3)$$

\noindent
At $P=2^{-2/3}P_1\simeq 0.6\,P_1$, $\eta^b_\gamma$
has a maximum, $\eta^b_\gamma= {3/8}$. In this case the total
luminosity is comparable with the rate of rotational energy loss.
The luminosities of all known $\gamma$-ray pulsars
can be explained by the modified polar-gap model.

The modified model is valid only if both
$B_{_S}>0.1B_{\rm cr}$ and $P_2<P<P_1$, where $P_2\simeq 0.07
(T_{_S}/10^6\,{\rm K})^{4/11}(B_{_S}/0.1B_{\rm cr})^{2/11}$ s.
For most of the $\gamma$-ray pulsars in Table~1,
such a strong magnetic field in the polar gap is suggested by the
surface dipolar component inferred from the spin down. For two of
the $\gamma$-ray pulsars (PSR 1055-52 and Geminga) the dipolar
estimate
is slightly below the required value. Nevertheless, it is plausible
that the field in the polar cap is strong for bound-pair formation
in these two cases provided one invokes higher-order multipolar
components, or an off-centered dipole.

The Crab-like pulsars (PSR 0531+21 and PSR 0540--69) have periods
shorter than $P_2$, and the high-frequency radiation
from these pulsars cannot be explained in terms of the
modified model. For these two Crab-like pulsars,
the outer-gap model of Cheng et al. (1986a) seems satisfactory
(Ulmer et al. 1994).
Moreover, the $\gamma$-ray emission from the Crab pulsar may also
be explained in terms of the slot gap model of Arons (1983). However,
the slot gaps are an effective source of energy for non-thermal
radiation only for dipole-like magnetic fields.

Some of the particles created near the top edge of the polar gap are
stopped by the field $E_\parallel$ and then accelerated back to the
star. By bombarding the pulsar surface, the reversed particles
heat it locally in the polar cap region. In the modified polar-gap
model the polar-cap temperature is about $T_e$ or $T_i$, depending
on whether electrons or ions escape from the surface. Both
$T_e$ and $T_i$ are a simple function of either the work
function for electrons or the cohesive energy for ions (e.g.,
Usov \& Melrose 1995).
An interesting implication of the modified model for $\gamma$-ray
pulsars is that, in principle, information on the binding of
particles to the polar cap and of the $B$-field strength in the
polar caps may be deduced from X-ray observations.

\subsection{Non-thermal high-frequency radiation}

The primary particles accelerated in the polar gap move away from
the pulsar and generate $\gamma$-rays. Some part of
these $\gamma$-rays is absorbed in the pulsar magnetic field
creating secondary pairs. The secondaries can repeat this process,
which leads to the development of cascades.
Gamma-ray emission from such cascades has been studied by Monte
Carlo simulations. In these simulations, a pair cascade was
initiated by either Compton-scattered photons (Dermer
\& Sturner 1994) or curvature photons (Daugherty \& Harding 1994).
In both of these cases the $\gamma$-ray spectra of pulsars were
fitted
fairly well. If the rotation and magnetic axes are nearly aligned,
both broad single-peaked and sharp double-peaked pulse profiles
with $\sim 0.4-0.5$ phase separation are formed,
in agreement with observations of $\gamma$-ray pulsar pulse profiles.

If the pulsar magnetic field is nearly orthogonal dipole,
the polar cascade models have difficulty explaining the interpulse
$\gamma$-ray emission of pulsars. This difficulty might be overcome
by taking into account that in the polar-gap model
the cascades in the neutron star vicinity
may not be the only source of powerful high-frequency emission.
Plasma instabilities may be developed in the outflowing plasma.
For example, the cyclotron instability may be developed near the
light cylinder of pulsars (Machabeli \& Usov 1979). This instability
leads to pitch-angles of the plasma particles. As a result,
synchrotron radiation has to be generated from
the region of development of the cyclotron instability.

\section{ Conclusions and discussion}

Many processes, such as generation of electric fields, particle
acceleration,
generation of $\gamma$-rays and pair creation, which are relevant to
the transformation of the rotational energy of the neutron star into
high-frequency emission are considered fairly well.

Using available data on high-frequency radiation of $\gamma$-ray
pulsars, some conclusions about validity of the polar-gap models for
$\gamma$-ray pulsars may be done by now. For example, the Ruderman
\& Sutherland (1975) model in which there is no
ejection of particles from the stellar surface is ruled out for the
$\gamma$-ray pulsars. Indeed, in this model the polar gaps are
symmetric and the energy flux into the pulsar magnetosphere is equal
to the energy flux into the polar caps. Since practically
all energy flux into the polar caps is reradiated as  X-ray emission
but only a part of the energy carried by
relativistic particles into the pulsar
magnetosphere may be radiated in the form of $\gamma$-rays, the
pulsar  $\gamma$-ray luminosity from a polar cap accelerator alone
in this model cannot be more
than its thermal X-ray luminosity. This is in  contradiction with
recent observations.

Most probably, the $\gamma$-ray emission was observed until
recently only from peculiar pulsars from which the $\gamma$-ray
flux is anomalously high. Such a $\gamma$-ray flux
amplification may be because either the rotation and
magnetic axes are nearly aligned or the surface magnetic field is very
high. Besides, the $\gamma$-ray emission may be amplified by the
outer gap action if the pulsar period is small enough.
Maybe, PSR B0656+14 for which the $\gamma$-ray luminosity is small
(Ramanamurthly {\it et al.} 1996 ) is the first pulsar with
conventional
polar gaps as a source of the energy for the pulsar $\gamma$-ray
emission (cf., Harding {\it et. al.} 1993).


\end{document}